\documentclass{article}
\usepackage{emulateapj}
\usepackage{amsmath}
\usepackage{amssym}
\usepackage{psfig}
\usepackage{times}

\input{epsf}

\def\dd{{ d}}

\def\ob{{\it\bf d}}
\def\nn{{\it\bf n}}
\def\mumin{\mu_{\rm min}}
\def\mumax{\mu_{\rm max}}
\def\Fmin{F_{\rm min}}
\def\Fmax{F_{\rm max}}
\def\Am{A_{\rm max}}
\def\Fobs{F_{\rm obs}}
\def\dg{{\rm o}}
\def\be{\beta}

\newbox\grsign \setbox\grsign=\hbox{$>$} \newdimen\grdimen \grdimen=\ht\grsign
\newbox\simlessbox \newbox\simgreatbox \newbox\simpropbox
\setbox\simgreatbox=\hbox{\raise.5ex\hbox{$>$}\llap
     {\lower.5ex\hbox{$\sim$}}}\ht1=\grdimen\dp1=0pt
\setbox\simlessbox=\hbox{\raise.5ex\hbox{$<$}\llap
     {\lower.5ex\hbox{$\sim$}}}\ht2=\grdimen\dp2=0pt
\setbox\simpropbox=\hbox{\raise.5ex\hbox{$\propto$}\llap
     {\lower.5ex\hbox{$\sim$}}}\ht2=\grdimen\dp2=0pt

\lefthead{BELOBORODOV}
\righthead{BENDING OF LIGHT NEAR COMPACT OBJECTS}

\begin{document}

\title{Gravitational bending of light near compact objects}

\author{Andrei M. Beloborodov\altaffilmark{1}} 
\affil{Canadian Institute for Theoretical Astrophysics,
University of Toronto, 60 St. George Street, Toronto, 
ON M5S 3H8, Canada } 
\altaffiltext{1}{Also at Stockholm Observatory, 
                               SE-106 91 Stockholm, Sweden \\
and Astro-Space Center of LPI, Profsojuznaja 84/32, Moscow 117810, Russia}

\begin{abstract}
A photon emitted near a compact object at an angle $\alpha$ with respect 
to the radial direction escapes to infinity at a different angle 
$\psi>\alpha$. This bending of light is caused by a strong gravitational
field. We show that, in a Schwarzschild metric, the effect is described by 
$1-\cos\alpha=(1-\cos\psi)(1-r_g/R)$ where $R/r_g$ is the emission radius 
in Schwarzschild units. The formula is approximate and it applies at 
$R\geq 2r_g$ only, however at these radii it has amazing accuracy, fully 
sufficient in many applications. As one application we develop a new 
formulation for the light bending effects in pulsars. It reveals the simple 
character of these effects and gives their quantitative description with 
practically no losses of accuracy (for the typical radius of a neutron star 
$R=3r_g$ the error is  1\%). The visible fraction of a star surface is shown 
to be $S_v/4\pi R^2=[2(1-r_g/R)]^{-1}$ which is $3/4$ for $R=3r_g$. 
The instantaneous flux of a pulsar comes from one or two antipodal polar 
caps that rotate in the visible zone. The pulse produced by one blackbody 
cap is found to be sinusoidal (light bending impacts the pulse amplitude but 
not its shape). When both caps are visible, the pulse shows a plateau: the 
variable parts of the antipodal emissions precisely cancel each other. 
The pulsed fraction of blackbody emission with antipodal symmetry has an 
upper limit $\Am=(R-2r_g)/(R+2r_g)$. Pulsars with $A>\Am$ must be asymmetric.
\end{abstract}




\bigskip

\medskip

\section{Introduction}

Light bending by gravity is a classical effect of general relativity (e.g. 
Misner, Thorne, \& Wheeler 1973). In a spherically symmetric gravitational 
field, the exact bending angle is given by an elliptic integral (see Appendix). 
The effect plays important role in astrophysics; e.g. weak bending gives rise 
to the gravitational lensing phenomenon. Strong bending occurs near compact 
objects (black holes and neutron stars) and crucially affects their emission.

In this Letter we give a simple formula for the bending angle that replaces the 
elliptic integral with high accuracy at radii $R\geq 2r_g$. Here $r_g=2GM/c^2$ 
is the Schwarzschild radius of the gravitating spherical object of mass $M$. 
We apply this formula to neutron stars and develop a new simple formalism for
the light bending effects in pulsars. We illustrate with pulsars that have two 
point-like antipodal spots emitting thermal radiation, and compare the results 
with the exact theory (Pechenick, Ftaclas, \& Cohen 1983). The advantage of our 
formalism is that it gives a clear understanding of the bending effects and 
describes them analytically with almost no losses of accuracy as long as stars 
with $R>2r_g$ are considered. The exact theory well describes stars with any 
$R$, including $R<2r_g$ where the bending angle $\beta$ can exceed $90^\dg$, 
however it requires a complicated numerical treatment. Standard models of 
neutron stars predict $R>2r_g$ with a typical $R\approx 3r_g$ (e.g. Shapiro 
\& Teukolsky 1983). Our formalism then works with $\sim 1\%$ accuracy and 
should be sufficient in many practical problems. It should also be good for 
accretion disks around Schwarzschild black holes: the disk inner edge is at 
$3r_g$ and the bulk of emitted radiation does not enter the region $R<2r_g$.


\section{The cosine relation}

Let an emitter be located at point $E$ at a radius $R$ (Fig.~1). 
A photon emitted along the radial direction escapes radially.
A photon emitted at some angle $\alpha$ with 
respect to radius ($\alpha$ is measured by the local observer at $E$) 
escapes along a bent trajectory. We will describe the trajectory
in polar coordinates $(r,\psi)$ where $\psi=0$ for the escape direction. 
The standard way of computing $\alpha(\psi)$ is given in Appendix. 
Instead one can use the approximate relation
\begin{equation}
\label{eq:cos}
    1-\cos\alpha=(1-\cos\psi)\left(1-\frac{r_g}{R}\right).
\end{equation}
The accuracy of relation~(\ref{eq:cos}) is remarkably high. It gives 
the bending angle $\beta=\psi-\alpha$ with a small fractional error 
$e=\delta\be/\be$ shown in Figure~2. For example for $R=3r_g$ the maximum 
error is $e_{\rm max}=3$\% (at $\alpha=90^\dg$) and $e<1$\% for 
$\alpha<75^\dg$. High accuracy is maintained for $R>2r_g$. For $R<2r_g$ the 
approximation is not applicable: then $\cos\alpha=0$ would correspond to 
$\cos\psi<-1$. 
We therefore limit our consideration to radiation escaping from 
$R>2r_g$ where $\be<90^\dg$. 

The high accuracy of equation~(\ref{eq:cos}) is a striking property of the 
Schwarzschild spacetime. This functional shape cannot be understood as a 
linear expansion in $r_g/R$ (that expansion is given in Appendix and its 
accuracy is much worse). Equation~(\ref{eq:cos}) can be understood as
follows. Let us consider $x=1-\cos\alpha$ as a small parameter and 
expand $y=1-\cos\psi$ in terms of $x^k$. After some algebra we find

\begin{eqnarray}
\label{eq:yx}
\nonumber
  y=\frac{x}{1-u}-u^2\left[\frac{1}{112}\left(\frac{x}{1-u}\right)^3
  \right. \\ \left.
  +\frac{1}{224}\left(\frac{5}{3}-u\right)\left(\frac{x}{1-u}\right)^4
   +O(x^5)\right],
\end{eqnarray}  
where $u=r_g/R$. The linear term is just equation~(\ref{eq:cos}); it describes 
the standard solid angle transformation 
$\dd\Omega_0=2\pi\dd x\rightarrow\dd\Omega=2\pi\dd y$ 
along the radial ray. The interesting property of equation~(\ref{eq:yx}) is that
the $x^2$-term vanishes and the higher order terms have small coefficients. 
It explains the extremely high accuracy of equation~(\ref{eq:cos})
at small $u$ and/or $x$ (Fig.~1). Only when both $x\rightarrow 1$ 
and $u\rightarrow 0.5$ there is a sizable correction to the linear term
and here $e$ peaks sharply. The linear expansion in $u$ (also derived
in Appendix) reads $y=x(1+u)+O(u^2)$.

Though the cosine relation (\ref{eq:cos}) is sufficient for our subsequent 
discussion, note that it also gives a simple description for the shape of 
a photon trajectory with a given impact parameter $b$ (Fig.~1). 
Equation~(\ref{eq:cos}) can be written at any point $(r,\psi)$ along the 
trajectory if $\alpha$ measures the angle between the photon velocity and 
the local radial direction. Combining~(\ref{eq:cos}) with (\ref{eq:alpha})
we find $r^2=b^2[\sin^2\psi+(1-\cos\psi)^2r_g/r]^{-1}$ and
\begin{eqnarray}
\label{eq:traject}
  r(\psi)=\left[\frac{r_g^2(1-\cos\psi)^2}{4(1+\cos\psi)^2}
  +\frac{b^2}{\sin^2\psi}\right]^{1/2} -\frac{r_g(1-\cos\psi)}{2(1+\cos\psi)}. 
  \;\;
\end{eqnarray}
Trajectories~(\ref{eq:traject}) are very close to the exact ones.
The fractional error of $r$ at a given $\psi$ is 
$\delta r/r<4\times 10^{-3}$ as long as $r>2r_g$.


\section{The visible surface of a star and the observed flux}

Consider a spherical compact star viewed by an observer at distance $D\gg R$.
Choose spherical coordinates $(R,\psi,\phi)$ with the polar axis $\psi=0$
directed toward the observer. What is the observed flux $\dd F$ from a surface 
element $\dd S$ of the star? First of all one should decide whether $\dd S$ is 
visible.  In a flat spacetime the visibility condition would be simply 
$\mu=\cos\psi>0$ where $\psi$ is the polar angle of $\dd S$. The star gravity 
creates curvature that allows the observer to see regions with negative $\mu$
down to a critical $\mu_v$ that defines the dark side of the star. The 
observed light from the circle $\mu=\mu_v$ is emitted tangentially to the 
star ($\alpha=90^\dg$). Exact calculations of $\alpha(\psi)$ via the elliptic 
integral yield tabulated $\mu_v$, e.g. $\mu_v=-0.886$, $-0.633$, $-0.484$ 
for $R/r_g=2$, 2.5, 3. From equation~(\ref{eq:cos}) we find immediately
the approximate $\mu_v=-r_g/(R-r_g)$. The visible fraction of the surface
is $S_v/4\pi R^2=(1-\mu_v)/2=[2(1-r_g/R)]^{-1}$, e.g. we get $0.75$ for
$R=3r_g$ while the exact value is 0.742.

Appendix gives the standard formula for the observed flux from a visible 
$\dd S$ (eq.~\ref{eq:flux}). It requires the knowledge of $\alpha(\psi)$ and 
the exact theory leads to a complicated numerical treatment of the problem 
(Pechenick et al. 1983). Instead we use the approximate 
equation~(\ref{eq:cos}); then equation~(\ref{eq:flux}) reads
\begin{eqnarray}
\label{eq:F}
  \dd F=\left(1-\frac{r_g}{R}\right)^2I_0(\alpha)\cos\alpha\frac{\dd S}{D^2},
\end{eqnarray}
where $\cos\alpha=\mu(1-r_g/R)+r_g/R$.
Equation~(\ref{eq:F}) is surprisingly simple. Apart from the redshift factor 
$(1-r_g/R)^2$ the element $\dd S$ emits as if the problem were Newtonian
with $\psi$ replaced by $\alpha$. 

\centerline{
\epsfxsize=10.7cm {\epsfbox{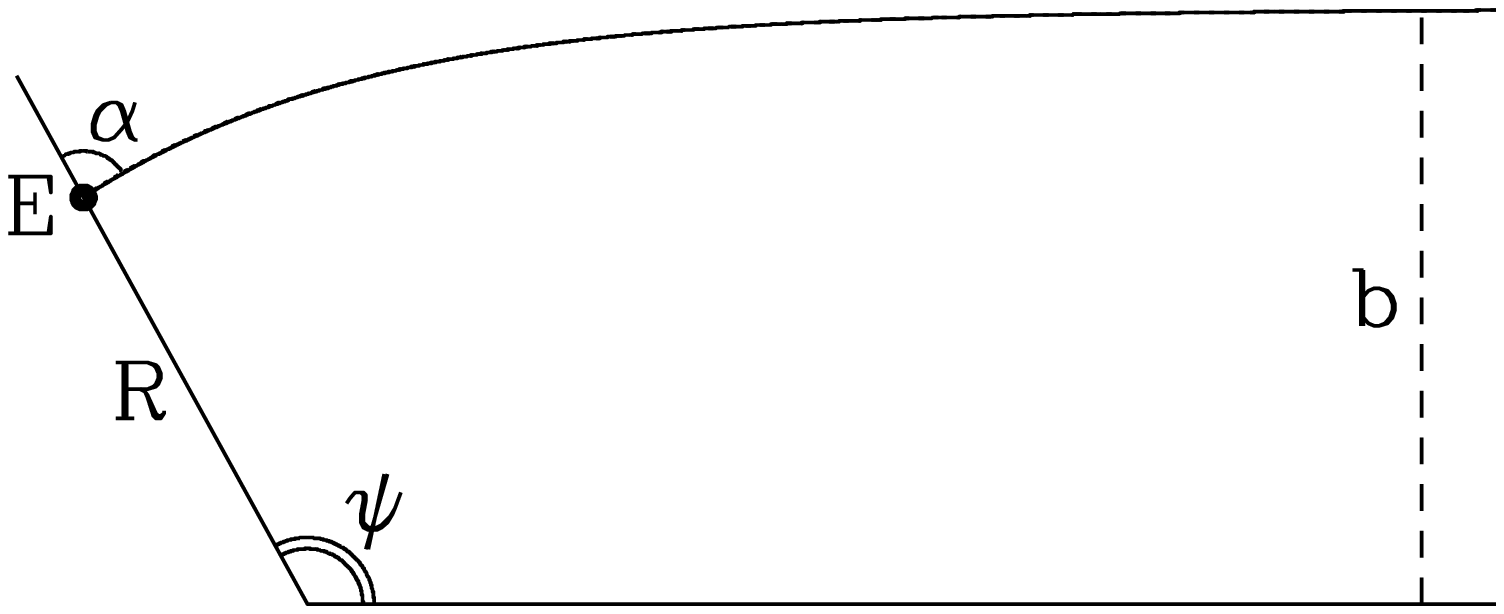}}
}
\bigskip
\figcaption{Example photon trajectory. The emission point $E$ is at
$R=2.5r_g$ and the emission angle is $\cos\alpha=0.1$; the corresponding
impact parameter $b=3.21r_g$. In this case
$\cos\psi=-0.482$ and the bending angle $\be=\psi-\alpha\approx 35^\dg$.
The approximate trajectory (eq.~\ref{eq:traject}) is also plotted here and
it is not distinguishable from the exact trajectory.
\label{fig1}}
\bigskip
\centerline{
\epsfxsize=9.5cm {\epsfbox{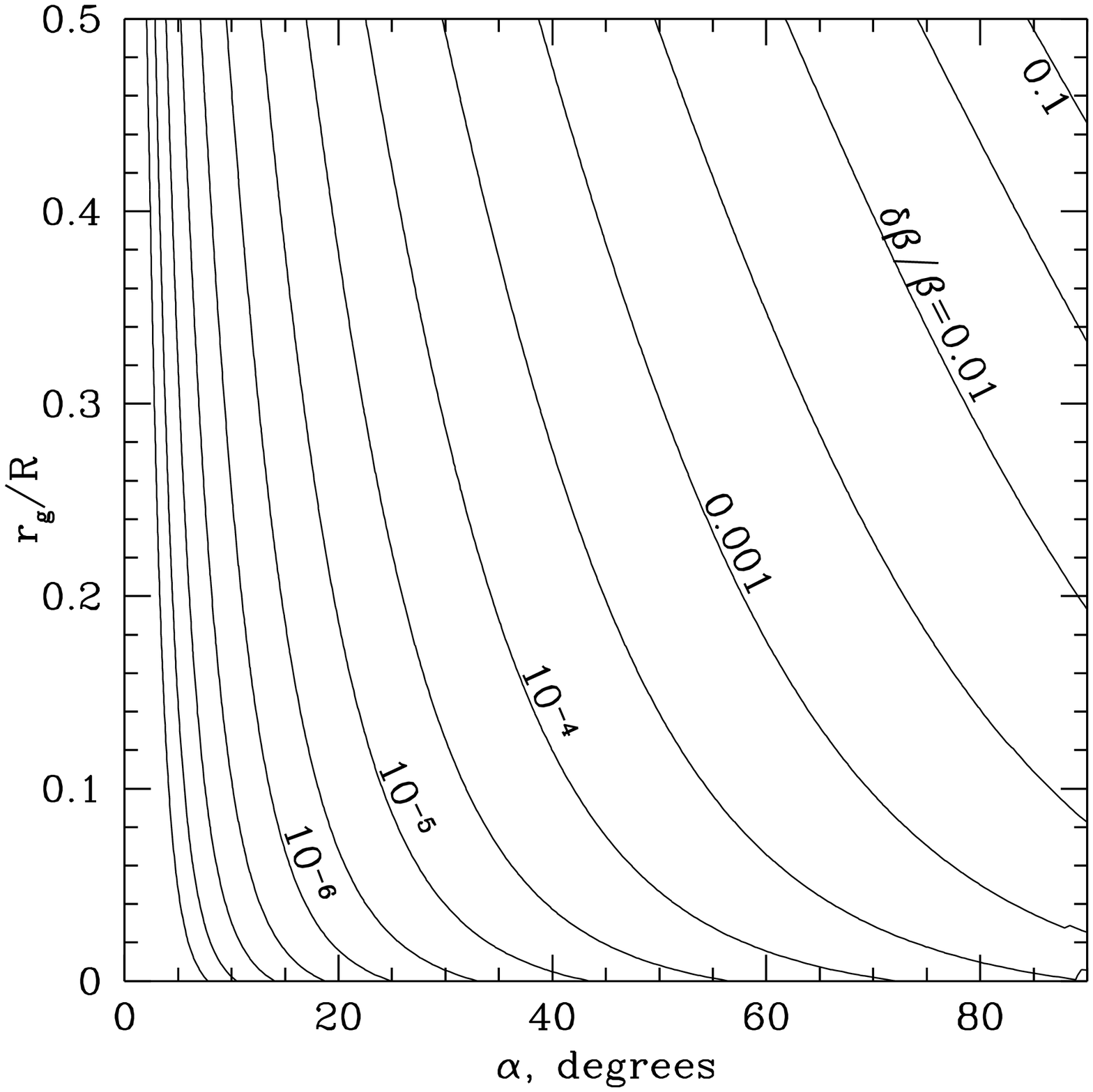}}
}
\bigskip
\figcaption{Accuracy of the cosine relation~(\ref{eq:cos}). This relation
gives the bendig angle $\beta=\psi-\alpha$ with an error
$e=\delta\beta/\beta$ which depends on the emission radius $R/r_g$ and the
emission angle $\alpha$. Here the contours of $e=const$ (with logarithmic
step 0.5) are plotted on the $r_g/R-\alpha$ plane.
\label{fig2}}
\medskip

\bigskip

\bigskip

When the local intensity of radiation $I_0(\alpha)$ is known for all $\dd S$,
equation~(\ref{eq:F}) can be integrated over the visible surface $\mu>\mu_v$
to get the total observed flux. For example, suppose a star has a homogeneous 
hot spot of arbitrary size and shape that emits thermal radiation 
[$I_0(\alpha)=const$]. Denote the spot area by $s=\int\dd S$ and the 
corresponding projected area by $s_\perp=\int\mu\dd S$ (if only a part of the 
spot is in the visible zone $\mu>\mu_v$ the integrals are taken over this part). 
Then the observed flux is $F=(I_0/D^2)(1-r_g/R)^2[s_\perp(1-r_g/R)+sr_g/R]$.
Note that $s_\perp<0$ when the spot is in the region $\mu_v<\mu<0$.

\newpage


\section{Pulsars}

Consider a neutron star with two antipodal hot spots associated with polar 
caps of the star magnetic field. Assume here that the spot size is small 
compared to the star radius $R$. The star is spinning and its magnetic axis 
is inclined to the spin axis by an angle $\theta\leq 90^{\rm o}$. Therefore 
the spots periodically change their position and inclination with respect to 
a distant observer and the observed radiation pulsates.

The spot whose circle of rotation is closer to the observer will be called
``primary'' and the other symmetric spot ``antipodal''.
Denote the unit vector pointing toward the observer by $\ob$ and the angle 
between $\ob$ and the spin axis by $i$. To compute the radiation 
fluxes from the primary and antipodal spots one needs to know their inclinations: 
$\mu=\nn\cdot\ob$ and $\bar{\mu}=\bar{\nn}\cdot\ob$, where $\nn$ and 
$\bar{\nn}=-\nn$ are the spot normals.
When the star executes its rotation, $\mu$ varies 
periodically between $\mumin=\cos(i+\theta)$ and $\mumax=\cos(i-\theta)$,
\begin{eqnarray}
  \mu(t)=\sin\theta\sin i\cos\Omega t+\cos\theta\cos i. 
\end{eqnarray}
Here $\Omega$ is the angular velocity of the pulsar; $t=0$ is chosen
when $\mu=\mumax$. 

Consider the blackbody component of the pulsar emission; it has isotropic
intensity $I_0(\alpha)=const$. Using (\ref{eq:F}) we find 
the observed fluxes $F$ and $\bar{F}$ from the primary and antipodal spots 
(when they are visible) 
\begin{equation}
\label{eq:FF}
  \frac{F}{F_1}=\mu\left(1-\frac{r_g}{R}\right)+\frac{r_g}{R}, \quad
  \frac{\bar{F}}{F_1}=-\mu\left(1-\frac{r_g}{R}\right)+\frac{r_g}{R}.
\end{equation}
Here $F_1=(1-r_g/R)^2I_0s/D^2$, $s$ is the spot area, and 
point-like-spot approximation is adopted.
For the antipodal spot we substituted $\bar{\mu}=-\mu$.
The observer sees the primary spot when $\mu>\mu_v$ and the antipodal 
spot when $\bar{\mu}>\mu_v$ (\S~3).
Denote $\kappa=r_g/(R-r_g)=|\mu_v|$. Both spots are seen when 
$-\kappa<\mu<\kappa$ and then the observed flux is
\begin{equation}
\label{eq:plat} 
 \Fobs=F+\bar{F}=2\frac{r_g}{R}F_1=const {\rm \;\; !}
\end{equation}
Hence the blackbody pulse must display a plateau whenever both spots are 
in the visible zone. We emphasize the high accuracy of equations~(\ref{eq:FF}) 
and (\ref{eq:plat}), though the exact theory does not give the ideal plateau. 
The fractional error $\delta$ in $\Fobs$ is maximal when $\mu=\mu_v$, e.g. 
$\delta_{\rm max}\approx 1.3$\% for $R=3r_g$. For $R=2r_g$ our formalism 
predicts that the whole surface is visible and $\Fobs$ is given by 
equation~(\ref{eq:plat}) at any $\theta,i,\mu$; then $\delta_{\rm max}\approx 6$\%. 

At any time $\Fobs$ takes one of three values: $F$, $\bar{F}$, or $F+\bar{F}$. 
A pulsar shows the plateau~(\ref{eq:plat}) if $\cos(i+\theta)<\kappa$.
A pulsar with arbitrary $i$ and $\theta$ belongs to one of four classes 
described in Figures~3 and 4. It is easy to compute the pulsed fraction 
$A=(\Fmax-\Fmin)/(\Fmax+\Fmin)$ for each class,
\begin{eqnarray}
\label{eq:A}
A=\left\{ \begin{array}{ll}
    (\mumax-\mumin)/(\mumax+\mumin+2\kappa) & {\rm class\; I}\\
    (\mumax-\kappa)/(\mumax+3\kappa) & {\rm class\; II,III\;\;\;} \\
                 0              & {\rm class\; IV}
          \end{array}
       \right.
\end{eqnarray}
$A$ reaches its maximum $\Am$ if the pulsar has $\mumax=1$ and $\mumin<\kappa$.
This maximum is $\Am=(R-2r_g)/(R+2r_g)$.

\centerline{
\epsfxsize=8.8cm {\epsfbox{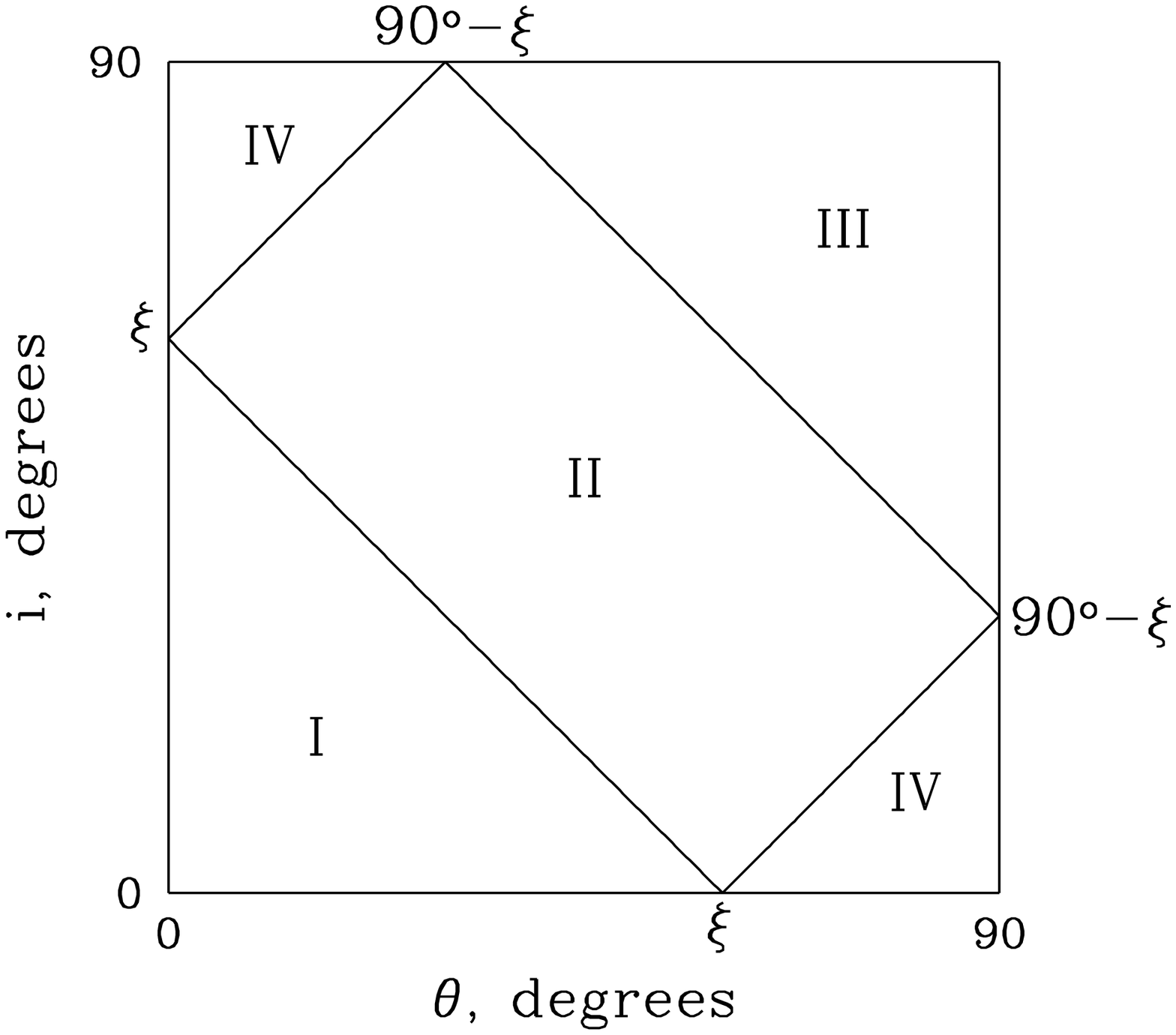}}
}
\figcaption{Location of classes I--IV on the $i-\theta$ plane.
The angle $\xi$ is defined so that $\cos\xi=r_g/(R-r_g)$, e.g. $\xi=60^\dg$
for $R=3r_g$. The classes are determined by the relative positions of
$\mumin=\cos(i+\theta)$, $\mumax=\cos(i-\theta)$, $\kappa\equiv r_g/(R-r_g)$,
and $-\kappa$. {\bf Class~I}: $\mumin>\kappa$. The antipodal spot is never
seen in such a pulsar (and the primary spot is visible all the time).
The observed blackbody pulse has a perfect sinusoidal shape. {\bf Class~II}:
$-\kappa<\mumin<\kappa<\mumax$. The primary spot is seen all the
time and the antipodal spot also appears for some time. The pulse
has a sinusoidal profile $F(t)$ interrupted by the plateau~(\ref{eq:plat}).
{\bf Class~III}: $\mumin<-\kappa$. The primary spot is not visible for a
fraction of period (and then only the antipodal spot is seen).
Such a pulsar shows a primary sinusoidal profile $F(t)$ interrupted by
the plateau~(\ref{eq:plat}) and the plateau is interrupted by a weaker
sinusoidal sub-pulse $\bar{F}(t)$ from the antipodal spot.
The sub-pulse occurs when only the antipodal spot is in the visible zone.
{\bf Class~IV}: $-\kappa<\mumin,\mumax<\kappa$. Both spots are seen at any
time. Then the observed blackbody flux is constant (eq.~\ref{eq:plat}).
Only anisotropic emission with $I_0(\alpha)\neq const$ can contribute to
pulsations.
}
\medskip

\centerline{
\epsfxsize=9.0cm {\epsfbox{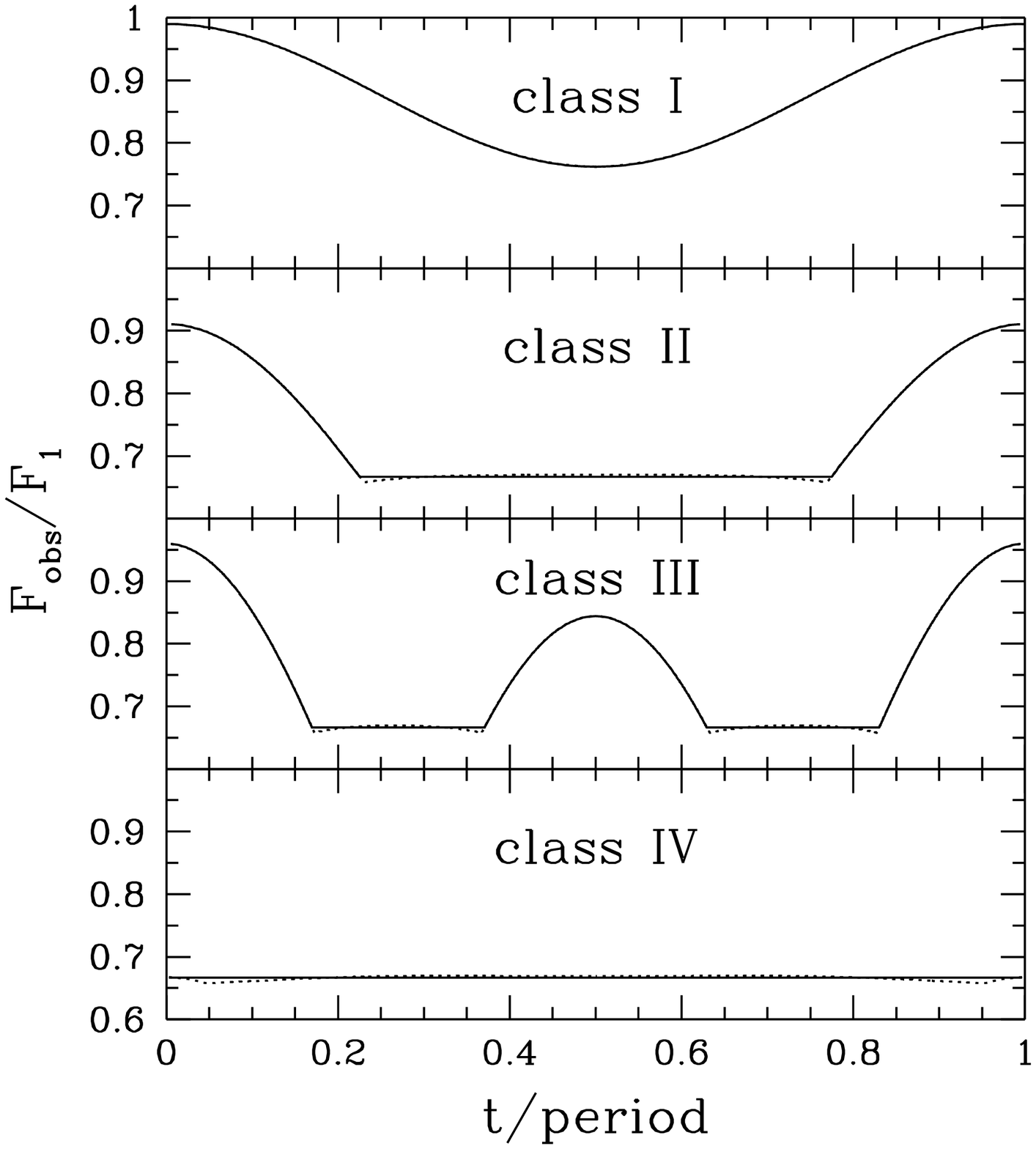}}
}
\figcaption{Blackbody pulse profiles for classes I--IV defined in Fig.~3.
$R=3r_g$ is chosen and the flux is normalized to $F_1$ (the maximum possible
flux that is observed when the primary spot is viewed face on).
Class I is exemplified with $\theta=20^\dg$ and $i=30^\dg$, class II
with $\theta=30^\dg$ and $i=60^\dg$, class III with $\theta=60^\dg$ and
$i=80^\dg$, and class IV with $\theta=20^\dg$ and $i=80^\dg$.
The approximate model developed in this paper is shown by solid curves
and the exact model is shown by dotted curves. The plateau has the same
level in all cases: $F_p/F_1=2r_g/R=2/3$.
}
\bigskip


\section{Conclusions}

A new formulation for the light bending effect is given in this Letter.
It makes use of an approximate simple formula that describes photon 
trajectories with remarkable accuracy (eq. 1).

This formulation is applied to pulsars where light bending is 
known to be a crucial effect. The developed analytical theory
replaces the previously used complicated numerical treatment with 
high accuracy, sufficient for practical purposes. For example, it immediately
gives the observed pulse profile for a rotating neutron star with 
two antipodal hot spots (eqs.~\ref{eq:FF} and \ref{eq:plat}). 
These analytical expressions highlight the effects of light bending
in pulsars:
(i) If the antipodal spot is never seen, the blackbody pulse is perfectly
sinusoidal. The bending affects the pulsed fraction $A$ only (eq.~\ref{eq:A}). 
(ii) The visible surface of the star is significantly enhanced, 
$S_v/4\pi R^2=[2(1-r_g/R)]^{-1}$, and both spots may be seen for a fraction 
of period or all the time. Then the blackbody pulse shows a plateau 
$\Fobs=F_p=(2r_g/R)F_1$ where $F_1$ is the flux from one spot viewed face on. 
(iii) The observed flux cannot exceed $F_1$ and it cannot be smaller 
than $F_p$. The latter is a purely relativistic effect: $F_p=0$ in the
Newtonian limit $r_g/R\rightarrow 0$. The maximum pulsed fraction of a pulsar 
with antipodal symmetry is $\Am=(F_1-F_p)/(F_1+F_p)=(R-2r_g)/(R+2r_g)$. This 
is a severe constraint, e.g. $\Am=1/5$ for $R=3r_g$ instead of Newtonian 
$\Am=1$.  

Within a few percent the results agree with the previous numerical works. 
The new formalism, however, is much simpler in use. It gives a clear description 
of the pulse profiles for given $i$, $\theta$, and $R/r_g$ and makes it easier 
to compare the theory with observations and constrain pulsar parameters.
We focused in this Letter on pulsars with small hot spots with opening 
angles $\delta\theta<10^\dg$. The extension to big spots is straightforward 
(\S~3). Note here that the plateau is predicted when both spots are fully 
visible and the upper limit $\Am$ will remain valid for blackbody emission 
with antipodal symmetry.

Caps of real pulsars (both radio and X-ray) do not emit as a blackbody
and their $I_0(\alpha)$ need not be isotropic. In particular: 
(1) X-ray pulsars show Comptonized power-law spectra, sometimes along with a 
clear blackbody component (as in SAX~J1808.4-3658). 
Light bending affects the two components differently: the blackbody pulse
from one spot remains sinusoidal while the Comptonized pulse is distorted. 
This distortion is described by equation~(\ref{eq:F}) and it depends on 
the anisotropic intensity $I_0(\alpha)$ of the Comptonized emission.
(2) For pulsars of high spin (with ms periods) the Doppler effect changes 
$I_0(\alpha)$. It is easily added to the model by the corresponding 
transformation of $I_0$ from the spot rest frame to the local non-rotating 
frame. (3) A strong magnetic field $B\sim 10^{12}$~G leads to anisotropic 
emissivity with a preferential direction along the field.
In any case, calculations of $I_0(\alpha)$ separate from the light bending 
effects. Radiation that propagates freely through the pulsar magnetosphere 
is described by equations~(\ref{eq:cos}) and (\ref{eq:F}) for any $I_0(\alpha)$.



\begin{center}
  APPENDIX: LIGHT PROPAGATION IN A SCHWARZSCHILD METRIC
\end{center}

For a given photon trajectory let us choose Schwarzschild coordinates 
$x^k=(t,r,\theta,\psi)$ so that the trajectory is in the plane $\theta=90^\dg$. 
Let $u^k=\dd x^k/\dd\lambda$ be the 4-velocity of the photon ($\lambda$ is 
an affine parameter, $\lambda=r$ at $r\rightarrow\infty$). The trajectory has 
integrals $u_t=1$ and $u_\psi=b$. 
From $u^iu_i=0$ one finds $(u^r)^2=1-(b^2/r^2)(1-r_g/r)$. Let us choose
$\psi=0$ for the escape direction. Then at radius $R$ the photon has
\begin{equation}
\label{eq:int}
   \psi=\int_R^\infty\frac{-u^\psi}{u^r}\;\dd r
       =\int_R^\infty\frac{\dd r}{r^2}\left[\frac{1}{b^2}-\frac{1}{r^2}
        \left(1-\frac{r_g}{r}\right)\right]^{-1/2}.
\end{equation}
The emission angle $\alpha$ (between the emitted photon and the local 
radial direction) is $\tan\alpha=\sqrt{u^\psi u_\psi}/\sqrt{u^ru_r}$ which 
yields
\begin{equation}
\label{eq:alpha}
   \sin\alpha=\frac{b}{R}\sqrt{1-\frac{r_g}{R}}.
\end{equation}
Combining equations~(\ref{eq:int}) and (\ref{eq:alpha}) one can compute
numerically the relation between $\psi$ and $\alpha$ for a given $R$. 
At small $u=r_g/R$ one can expand equation~(\ref{eq:int}) in $u$ and 
keep the liner term only. Using the equality 
$\int_0^{\sin\alpha}(\sin\alpha-z^3)(1-z^2)^{-3/2}\dd z=2(1-\cos\alpha)$
one gets $\psi=\alpha+u(1-\cos\alpha)/\sin\alpha+O(u^2)$.

Consider a star of radius $R$ with a surface radiation intensity 
$I_0(\alpha)$ which can vary over the surface. The observer location is 
given by a radius-vector ${\bf D}$ ($D\gg R$). A (visible) surface element 
$\dd S=R^2\dd\mu\,\dd\phi$ is observed at impact parameters $(b,b+\dd b)$ and 
subtends a solid angle $\dd\Omega=b\,\dd b\,\dd\phi/D^2$ on the observer sky.
Here $\mu=\cos\psi$ and $\phi$ is an azimuthal angle corresponding to 
rotation around ${\bf D}$; $b$ depends on $\mu$ only, not on $\phi$. 
The observed radiation flux from $\dd S$ is $\dd F=I\dd\Omega$ where 
$I=(1-r_g/R)^2I_0$ [$I$ and $I_0$ are related through $I/\nu^4=I_0/\nu_0^4$
and the frequency redshift $\nu/\nu_0=(1-r_g/R)^{1/2}$, see 
Misner et al. 1973]. Using equation~(\ref{eq:alpha}) one gets
\begin{equation}
\label{eq:flux}
   \dd F=\frac{Ib}{R^2}|\frac{\dd b}{\dd\mu}|\frac{\dd S}{D^2} 
   =\left(1-\frac{r_g}{R}\right)I_0(\alpha)\cos\alpha
     \frac{\dd\cos\alpha}{\dd\mu}\frac{\dd S}{D^2}.
\end{equation}




\begin{references}

\reference{}
Misner, C. W., Thorne, K. S., \& Wheeler, J. A. 1973, Gravitation,
(San Francisco: Freeman)

\reference{}
Pechenick, K. R., Ftaclas, C., \& Cohen, J. M. 1983, ApJ, 274, 846

\reference{}
Shapiro, S. L., \& Teukolsky, S. L. 1983, Black Holes, White Dwarfs,
and Neutron Stars, (New York: Wiley-Interscience)


\end{references}
\end{document}